\documentclass{aamed}
\usepackage{graphicx}

\begin{document}

\title {Microflares in accretion disks}
\author{V.\ Krishan\inst{1}
  \and S. Ramadurai\inst{2}$^,$\inst{3}
  \and  P.\ J. Wiita\inst{4}} 

\offprints{P.\ J.\ Wiita} 
       
\institute{Indian Institute of Astrophysics,
        Bangalore 560034, India
        (vinod@iiap.ernet.in)
\and 
Tata Institute of Fundamental Research, Theoretical Astrophysics Group,
Mumbai 400005, India 
\and
Institutt for Fysikk, NTNU -- Norwegian University of Science and Technology,
N7491, Trondheim, Norway (ramadura@phys.ntnu.no)
\and
Department of Physics \& Astronomy, Georgia State University,
Atlanta GA 30303, USA (wiita@chara.gsu.edu)}

\date{Received 20 July 2001 / Accepted 13 November 2002}

\abstract{
 We have investigated the phenomenon of explosive chromospheric evaporation 
from an accretion disk as a mechanism for fast variability in accreting sources such
as low mass X-ray binaries and active galactic nuclei.  This has
been done in
the context of advection dominated accretion flows, allowing both
high and low states to be considered.  This mechanism can in principle 
produce sub-millisecond timescales in binaries and sub-minute timescales
in active galaxies.  However, even considering the possibility that large 
numbers of these
microflares may be present simultaneously, the power
 emitted from these microflares probably amounts to only a small
fraction of the total X-ray luminosity.


}

\maketitle

\section {Introduction}

All models for X-ray binaries (XRBs) involve accretion onto compact central
objects.
Similarly, accretion onto supermassive black holes is the
dominant paradigm for the engine of active galactic nuclei (AGN).
Recently, advection dominated accretion flows (ADAFs) have been advocated
as a natural way of explaining the variety of states in which individual
binary
X-ray sources have been observed  (e.g.\ Chakrabarti \&
Titarchuk \cite{chakrabarti95};
Esin et al.\ \cite{esin97}, \cite{esin98}).
Rapid fluctuations  have been detected in many electromagnetic bands and
modeled in a wide variety of ways (e.g.\ Krishan \& Wiita \cite{krishan94}; 
Wagner \& Witzel \cite{wagner}; Tanaka \& Shibazaki \cite{tanaka}; Wiita \cite{wiita}).
In low-mass X-ray binaries, we are presumably dealing with
accretion onto black holes (BHs) with masses typically 5--15 $M_{\odot}$,
with luminosities of $L_{\rm X} \sim 10^{35-37}$ erg s$^{-1}$ in the
low/hard state, and with fast variations
detectable down to $0.01$ s (e.g.\ Esin et al.\ \cite{esin97}, \cite{esin98};  Tanaka \&
Shibazaki \cite{tanaka}; 
Trudolyubov et al.\ \cite{trudolyubov}).
Seyfert galaxies typically exhibit $L_{\rm X} \sim 10^{42-44}$ erg s$^{-1}$, 
with noticeable variations observed on timescales as short as
 minutes (e.g.\ 
Lawrence \& Papadakis \cite{lawrence}; Turner et al.\ \cite{turner}).

Many accretion models involve hot low-density coronae lying over
 high density, cooler accretion disks (e.g.\ Galeev et al.\ \cite{galeev};
 White \& Holt \cite{white}; Haardt et al.\ \cite{haardt}; Poutanen \cite{poutanen}),
and much of the resulting physics is not dissimilar to that in ADAF-type models.
Recent three-dimensional magnetohydrodynamical simulations of accretion 
disks indicate
that magnetic fields will be strengthened within the disk and that plasma
and fields will escape from the disk into a strongly magnetized
corona (Miller \& Stone \cite{miller}).
This situation is reminiscent of the conditions prevailing in the
solar atmosphere, and therefore, our understanding of solar flare
related processes may
well be carried over to accreting sources such as 
XRBs and AGN, which exhibit rapid variations in luminosity
 (e.g.\ Kuijpers \cite{kuijpers}; di Matteo \cite{dimatteo}; Krishan et al.\ \cite{krishan00}). 
In particular, the phenomenon of
chromospheric evaporation caused by the nonthermal electron heating
is often invoked to explain the rapid rise of the soft X-ray emission
during flares and the blueshift of soft X-ray lines of
\ion{Ca}{xix} and \ion{Fe}{xxv} (e.g.\ Antonucci et al.\ \cite{antonucci}).  This explosive evaporation
has been studied using what is known as the exploding gasbag model 
(Fisher \cite{fisher}; Bastian et al.\ \cite{bastian}).  
 In this paper, we extend this model to conditions 
relevant to accretion flows in both LMXBs and AGNs, and make an approximate
evaluation of its possible contribution to X-ray emission and variability.

\section{Basic Model}

Figure 1 illustrates the situation we have in mind.  
We consider an ADAF 
allowing for variations in the accretion rate, ${\dot M}$, and thus in 
the transition radius, $r_t$, which is thereby
capable of reproducing the spectrum of X-ray binaries in both their
low/hard state and high/soft states (e.g.\ Esin et al.\ \cite{esin97}).  The
low (high)  state is
presumed to correspond to low (high) values of the accretion rate, 
normalized to the
Eddington accretion rate, ${\dot m} = {\dot M}/{\dot M}_{\rm Edd}$. 
For larger values of ${\dot m}$ the 
ADAF region shrinks ($r_t$ decreases)(e.g.\
Abramowicz et al. \cite{abramowicz}; Narayan et al.\ \cite{narayan97}); specific
ranges of $\dot m$ are capable of producing the quiescent,
intermediate and very-high states observed in some BH XRBs,
such as Nova Muscae 1991 (Esin et al.\ \cite{esin97}), Cygnus X-1, GRO J0422+32
and GRO J1719$-$24 (Esin et al.\ \cite{esin98}), but we shall not
be concerned with those details here.  The two temperature
plasma within the ADAF region extends into a corona above (at
least the inner portion of) the thin accretion disk.  We assume
that there is significant flare activity produced through
magnetic reconnection within this
corona (e.g.\ Kuijpers \cite{kuijpers}). Such releases of magnetic energy are presumably 
concentrated near the inner boundary of the disk where the densities and field
strengths within the disk are greatest (e.g.\ Chakrabarti \& D'Silva
\cite{chakrabarti94}), and thus we will consider parameters 
relevant to this region. 

\begin{figure}
  \resizebox{\hsize}{!}{\includegraphics{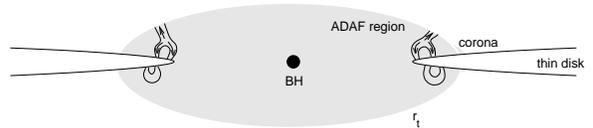}}
\caption{A schematic of an accretion flow in the low-state, where the transition
radius, $r_t$, is far from the black hole (BH).  The inner, two-temperature
plasma extends out to $r_t$, with a thin disk and a corona present beyond
that point.  Flares within the corona produce electron beams, some of which
impinge upon the cool disk and evaporate material from it into the corona.}
\label{fig1}
\end{figure}

Magnetic reconnection is now believed to be an essential component
of the solar flare process (see Lakhina \cite{lakhina} for a recent review).
Bi-directional electron beams (i.e., propagating both towards and away
from the Sun) are now understood to be a
signature of the magnetic reconnection process as revealed
through solar radio observations (e.g.\ Robinson \& Benz \cite{robinson}).
Here we investigate conditions under which the ``chromospheric
evaporation'' induced by downgoing electron beams (e.g.\ Fisher \cite{fisher};
Bastian et al.\ \cite{bastian}) could operate 
in ADAF type accretion disks giving rise to rapid variability
or microflares.
In analogy with the solar case, we anticipate that a large number of 
small flares will be present simultaneously (e.g.\ Parker \cite{parker}).
The evaporation of thin disk material can erode the inner portion
of the disk, thus changing $r_t$, and thereby possibly inducing additional
X-ray variability.

\section{Evaporation from accretion disks}

\subsection{Conditions appropriate to low states  for XRBs}

To illustrate this situation, we adopt simple standard
$\alpha$ disk models for the outer portion of the accretion
disk (e.g.\ Shakura \& Sunyaev \cite{shakura}, hereafter SS), but truncate them
at $r_t$, where the two-temperature advection dominated
portion of the flow is assumed to take over.  In order to incorporate
the explosive gasbag (EG) model for chromospheric evaporation
(e.g.\ Fisher \cite{fisher}) we need values for the density $n_{\rm ch}$
and the scale height, $z_0$ in that region. 
At this distance from the BH, the thin disk will be characterized
by gas pressure dominance over radiation pressure and
electron scattering opacity larger than free-free or other opacities,
 so (SS),
\begin{equation}
h = 2.3 \times 10^3 \alpha^{-1/10} m^{9/10} {\dot m}^{1/5} r^{21/20} {\rm cm},
\end{equation}
\begin{equation}
n_0 = 9.2 \times 10^{24} \alpha^{-7/10} m^{-7/10}{\dot m}^{2/5} r^{-33/20}{\rm cm}^{-3}.
\end{equation}
Here, $h$ is the disk half-thickness, $\alpha$ the viscosity parameter,
$m \equiv M_{\rm BH}/M_{\odot}$, $ {\dot m} = {\dot M}/(2.3 \times 10^{-9}M_{\odot}
{\rm yr}^{-1}m)$, $r = R/(2GM_{\rm BH}c^{-2})$, $n_0$ is the number density
of the disk in the equatorial plane, and $R$ is the physical distance from 
the BH, so the coefficients differ from those in Shakura \& Sunyaev (\cite{shakura}), which
uses different definitions of ${\dot m}$ and $r$.  We have dropped the 
relativistic correction terms which are negligible at the 
distances considered here.

These standard disk models do not provide precise values for the
density, $n_{\rm ch}$ and the scale height, $z_0$,
 in the photospheric/chromospheric 
regions of the disk.  Nonetheless, an estimate for the $z_0$ in the photosphere, 
where the density falls exponentially as $n \propto {\rm exp}(-z/z_0)$, is
given by $z_0 = r^3 k T(z_1)/(G M m_p z_1)$, where $m_p$ is the proton mass
and $z_1$ is given by the
location where the total optical depth equals unity (SS).
In that $z_1$ is only slightly greater than $z_0$ and the chromosphere begins
at a height of at least $2z_1$, we adopt values of 
$n_{\rm ch} =  1.0 \times 10^{-2}~ \chi~ n_0$ and
$z_0 =  0.3 ~\eta ~h$, where we expect $\chi$, and $\eta$ to be 
within an order of magnitude of unity.  

The microflares are likely to be concentrated in the transition region,
and we use parameters relevant for low-states of ADAF models.  This implies
that $\dot m < 1.3 \alpha^2$ (e.g.\ Esin et al.\ \cite{esin97}).  Most current estimates
for disk viscosity involve some form of the magnetorotational instability
and tend to produce $\alpha \le 0.1$ (e.g.\ Brandenburg \cite{brandenburg}; Hawley et al.\
\cite{hawley}), and
we thus assume typical values of $\alpha_{.1} = \alpha/0.1 \simeq 1$,
${\dot m}_{.01} = {\dot m}/0.01 \simeq 1$.  For these low states, $r_t \sim
10^2 - 10^4$, so we parameterize by taking $r_{100} = r_t/100$.  
We shall first focus on LMXBs, for which a typical BH mass of $10 M_{\odot}$
can be assumed, and thus use $m_{10} = M_{\rm BH}/(10 M_{\odot}) \simeq 1$.
Then we obtain the following estimates for the scale height and density
in the chromospheric regions
\begin{equation}
z_0 = 3.4 \times 10^5 \eta~ \alpha_{.1}^{-1/10} m_{10}^{9/10} {\dot m}_{.01}^{1/5} 
r_{100}^{21/20}~{\rm cm},
\end{equation}
\begin{equation}
n_{\rm ch} = 7.3 \times 10^{18} \chi~ \alpha_{.1}^{-7/10} m_{10}^{-7/10} 
{\dot m}_{.01}^{2/5} r_{100}^{-33/20}~{\rm cm}^{-3}.
\end{equation}

During a reconnection event,  electron beams form  in 
the current sheets; some of these will be directed downwards into the
disk atmosphere and heat it. Radiative losses must  be considered in
this situation.  In this region of the accretion disk the standard photospheric
temperature is given by
\begin{equation}
T_s = 1.4 \times 10^5 \alpha_{.1}^{1/75} m_{10}^{-19/75} {\dot m}_{.01}^{28/75}
r_{100}^{-141/150}~{\rm K}. 
\end{equation}
The radiative loss function, $\Lambda(T)$, is maximized at  temperatures
near $2 \times 10^5$K (Raymond et al.\ \cite{raymond}), which is 
marginally above the value just found for $T_s$.  Thus it is
likely that such temperatures will be found in
the transition region between the photosphere and the corona of the accretion
disk for $r_{100} > 1$, and this maximum loss rate can be achieved.  In order for this
evaporation mechanism to function, the minimum heating 
rate per particle, $Q$, must at least  
 balance the maximum cooling rate, and is found to be
$Q = \Lambda(T) n_{\rm ch} \simeq 7 \times 10^{-22} {\rm erg~ s}^{-1}
{\rm cm}^{-3}~n_{\rm ch}$ (Fisher \cite{fisher}), or
\begin{displaymath}
Q = 5.1 \times 10^{-3} \chi~ \alpha_{.1}^{-7/10}m_{10}^{-7/10}
{\dot m}_{.01}^{-2/5} r_{100}^{-33/20}{\rm erg~ s}^{-1}.
\end{displaymath}
If the photospheric temperature exceeds $2 \times 10^5$K then the cooling
rate drops rapidly to about $Q = 1.5 \times 10^{-22} n_{\rm ch}$ by
 $T_s = 5 \times 10^5$K, and then much more slowly until $T_s \simeq
3 \times 10^6$K (Raymond et al. \cite{raymond}).  Under these cicumstances
heating rates several times lower than that given above can suffice.

The energy deposited by the electron beam(s) heats the chromosphere, part of which
evaporates high into the corona. If the other
transport processes (e.g.\ conduction) are not immediately significant,  then the
evaporation can occur in an explosive manner and can be described
by the gasbag model (e.g.\ Fisher \cite{fisher}).
Using the standard equations for mass, momentum and energy conservation,
one can derive the time profiles of the plasma parameters in this gasbag.
Of key importance is the timescale for build up of maximum pressure and consequent 
rapid evaporation
of the plasma, which is found to be (Fisher \cite{fisher})
\begin{equation}
\tau = \Bigl(\frac{m_p Z_0^2}{\pi Q}\Bigr)^{1/3},
\end{equation}
where $m_p$ is the proton mass, and $Z_0$ is the initial thickness of
the evaporating plasma.  As shown by Fisher (\cite{fisher}), if this
EG model is to work at all, $Z_0$ must differ from $z_0$ only by
a logarithmic factor not much larger than unity.   For the circumstances 
considered here, conduction
is not important until times exceeding  $\sim 10\tau$, and
so the time $\tau$ represents the characteristic time 
over which fluctuations are produced by this process.  Of course, other processes
which we are not considering here may well play significant roles 
and could produce flares by different mechanisms
on different timescales.
Identifying $Z_0 = \zeta z_0$ with $\zeta > 1$,  we have the following approximate expression 
\begin{equation}
\tau = 2.3 \times 10^{-4} \zeta^{2/3} \eta^{2/3} \chi^{-1/3} \alpha_{.1}^{1/6} 
m_{10}^{5/6}  r_{100}^{5/4}~{\rm s}.
\end{equation}
Therefore, for XRBs, timescales of less than a millisecond are possible
for the explosive evaporation from the cool disk, even in the low state,
where $r_t$ is at least 100.

The minimum electron flux needed for balancing the radiative losses
can be approximated by
\begin{displaymath}
F \simeq Q \Bigl( \frac{6 N_c}{\delta -2} \Bigr),
\end{displaymath} 
where $N_c = 9.15 \times 10^{18}{\rm cm}^{-2}
(E_{c,10})^2$ (McClymont \& Canfield \cite{mcclymont}; Fisher et al.\ \cite{fisheretal}).
Here, $\delta$ is the power-law index for the non-thermal distribution of beam 
electrons (in the sense that $f(E)dE \propto E^{-\delta}dE$), and
the solar flare observations indicate that $\delta \simeq 4-5$ (McClymont \&
Canfield \cite{mcclymont}).  The total column density of beam electrons, $N_c$, is
determined from collisional processes and depends upon
the cut-off energy to which they can penetrate, which is typically
5--20 keV, so we use $E_{c,10} = E_c/10 {\rm keV} \sim 1$.
This electron energy flux $F$ implies densities in the beam several orders of 
magnitude
lower than the ambient density, $n_{\rm ch}$,  so the standard acceleration 
mechanisms
invoked in solar flares are adequate (e.g.\ Bastian et al.\ \cite{bastian}).
Explosive evaporation occurs for fluxes  greater than
this minimum value, and this EG process is not relevant if the fluxes are less
than this.

To obtain the total power emitted in a microflare we must estimate the
area over which this flux emerges.  Again, in analogy with the sun, we
approximate the size, $R$, of the flaring region to be of the order of the
scale height, $z_0$.   However, it is certainly possible that the flaring loop
could extend into the corona far enough so that the appropriate size is
actually that of the coronal scale height, which would be several times larger (e.g.\ McClymont \& Canfield \cite{mcclymont}).
Therefore, since nearly all of this power will be quickly radiated
in EUV and X-ray photons, we can obtain an estimate for the minimum total power in one of
these microflares as, $L_{f} = \pi R^2 F$; taking $\delta = 4$ we have,
\begin{equation}
L_{f}  \simeq   5.1 \times 10^{28}  \eta^2 \chi \alpha_{.1}^{-9/10} 
m_{10}^{11/10} {\dot m}_{0.01}^{-4/5}  r_{100}^{9/20} E_{c,10}^2 
\Bigl(\frac{R}{z_0}\Bigr)^2{\rm erg~s}^{-1}.
\end{equation}
The power in a single microflare $L_f$ is more than six orders
of magnitude smaller than the total power in the low state, using
our nominal values of $\eta \simeq \chi \simeq (R/z_0) \simeq 1$, 
so the sobriquet microflare is quite appropriate. For these low 
state conditions, ($r_t > 100$),
each of these nano/micro-flares covers an area very
small in comparison with the active portion  of the disk ($\sim 10^{-6}$).
It is essentially impossible to properly estimate the fractional area 
of the relevant part of
the accretion disk that will be subject to this process at any particular time.
However, it is certainly plausible to expect that no more than 10\% of the disk should
be so involved, and thus it is unlikely that $> 10^5$ of these explosive events
can be going on simultaneously.  Even with this optimistic assumption,
 the integrated output of all
these microflares is likely to remain small in comparison with the basal 
luminosity.  Only if most of the variables in Eq.\ (8) exceed their
nominal values could these microflares make a significant contribution
to the total power.  Probably the least certain factor in Eq.\ (8) is
the ratio $R/z_0$, and it is not implausible that this geometrical factor could 
approach 10; in that case, the individual microflare luminosities
 rise to above $10^{-6}$ 
of the observed total low state power.  Only in this optimistic situation is
it possible that the integrated flux from all such flares could amount
to a substantial fraction of the total emission.  However, for large values of
$R$,   we  expect there to be
fewer of these ``miniflares", in that each now covers a larger fraction of
the disk area.  Hence we do not expect the integrated micro/miniflare
power to rise as steeply as $(R/z_0)^2$.

\subsection{High/soft state conditions for XRBs}

Here we consider the situation where the accretion rate is high enough
so that the standard accretion flow penetrates much closer to the BH.
Then the relevant equations are those appropriate for the inner portion
of accretion disks, which are dominated by
radiation pressure (SS).  We note that for a high
state, ${\dot m} \sim 0.1$ and $r_t \sim 10$ are typical parameters;
 furthermore, the relativistic
corrections are no longer negligible, since we are dealing with
regions close to the BH.  With the reparameterizations ${\dot m}_{.1} = {\dot m}/0.1$
and $r_{10} = r/10$, defining $B \equiv [1 - (0.3/r_{10})^{1/2}]$, 
and still adopting $z_0 = 0.3 \eta h$ and $n_{ch} = 1.0 \times 10^{-2} \chi n_0$, 
we find:
\begin{equation}
z_0 = 7.4 \times 10^4  ~\eta  m_{10} {\dot m}_{.1} B ~{\rm cm},
\end{equation}
\begin{equation}
n_{\rm ch} = 4.4 \times 10^{20} \chi \alpha_{.1}^{-1} m_{10}^{-1} {\dot m}_{.1}^{-2} 
 r_{10}^{3/2} B^{-2}~{\rm cm}^{-3}. 
\end{equation}

Under these circumstances the photospheric temperature is given by
\begin{equation}
T_s = 6.6 \times 10^5 \alpha_{.1}^{1/5} m_{10}^{-1/5} {\dot m}_{.1}^{4/5} 
r_{10}^{-3/2} B^{4/5}~{\rm K}, 
\end{equation}
or $T_s \simeq 3.5 \times 10^5$K for $r_{10} = 1$, and all other quantities taking
on their new nominal values.  This somewhat exceeds the temperature at which maximum cooling
occurs.  It remains reasonable to utilize the value of $\Lambda$ corresponding to
this higher temperature to compute the heating rate needed to compensate for
cooling in the optically thin regime; for this temperature, $\Lambda$ is roughly half that
of the peak.

Proceeding as in the previous subsection we then obtain,
\begin{equation}
Q \simeq 1.5 \times 10^{-1} ~ \chi \alpha_{.1}^{-1} m_{10}^{-1} {\dot m}_{.1}^{-2}  
r_{10}^{3/2} B^{-2}{\rm erg~s}^{-1},
\end{equation}

\begin{equation}
\tau = 2.7 \times 10^{-5} \zeta^{2/3} \eta^{2/3} \chi^{-1/3} \alpha_{.1}^{1/3} 
 m_{10} {\dot m}_{.1}^{4/3} r_{10}^{-1/2} B^{4/3}~{\rm s},
\end{equation}
and
\begin{equation}
L_f = 7 \times 10^{28} \eta^2 \chi \alpha_{.1}^{-1} m_{10}~ r_{10}^{3/2} 
E_{c,10}^2 \Bigl(\frac{R}{z_0}\Bigr)^2 {\rm erg~s}^{-1}.
\end{equation} 
In that the total power for an XRB in the high state is 
$\sim 10^{37} {\rm erg~s}^{-1}$, we see that each of these micro(nano?)flares
 contributes very little
to the total power output.  Of course, as for the low state,  many such small
scale flares may contribute simultaneously;
however, in that the relevant disk area under the intense coronal region
is proportional to $r^2$ and is thus smaller than in the low state case,  the
integrated contribution is very likely to be negligible.

\subsection{Results for AGN}

Active galactic nuclei may also be described within the ambit of ADAF models,
particularly those with small output, and therefore probably low efficiency
(e.g.\ Lasota et al.\ \cite{lasota}; Narayan et al.\ \cite{narayan96}) so essentially 
everything developed above can still be applied.

The photospheric temperatures will be significantly lower, since they decline
with increasing black hole mass, but coronae
and transition regions should still be present, and microflares can still
evaporate disk material into the corona.  Appropriate values for all
dimensionless parameters are nearly the same as in the two
preceding subsections.  Therefore the timescales vary as $M^{5/6}$ for the
low state and as $M$ for the high state.  Roughly, then, these microflares
operate on 3--10,000 s timescales (for $M_{BH} \sim 10^7 - 10^8  M_{\odot})$.

Since $L_f \propto M^{11/10}$ for the low state and $L_f \propto M$ for the
high state, each of these microflares could contribute between $\sim 10^{34}-10^{36}$
erg s$^{-1}$, while the total powers are $\sim 10^{40}$ erg s$^{-1}$ for
low-state emission around a $10^7 M_{\odot}$ BH up through $\sim 10^{45}$  
erg s$^{-1}$ for
high-state emission near a $10^8$ solar mass black hole. 
Thus, even considering the possiblity that many of these microflares can erupt
 simultaneously, their
contribution to the total power still should be very small.
 
\section {Conclusions}

The mechanism of explosive chromospheric evaporation seems
to explain some aspects of solar X-ray variability, as well as 
blue-shifted soft X-ray lines from the Sun.  Since conditions
in accretion disk atmospheres probably have many similarities to the solar
case, we have explored this mechanism  in the context of accretion
disk models for XRBs and AGN.  In particular, we have focused on
the transition region in ADAF models as the most likely place for
these flares and microflares to be generated.  We have estimated
microflare timescales and powers based on the exploding gas bag
model for both low and high states
for both XRBs and AGN.  

We note that we have not proven that this
mechanism is necessarily relevant; we have merely shown that it
is plausible that it could operate in the conditions of accretion
disk atmospheres, as it probably does in the Sun.  Furthermore, 
it is very likely that other
flare mechanisms dominate the observed variability.
 For XRBs, the EG timescales range from sub-ms to
seconds, while for AGN, timescales from seconds to hours are expected,
with most of the relevant emission emerging up to $\sim 10$ keV.  Whereas
these characteristic times are in the range of the fastest observed
variability for both classes of sources, the powers in individual
microflares are extremely small.  However, we do expect a large
number of these flares could exist essentially simultaneously, 
particularly in the low states.
Even so, the integrated emission from these microflares 
 is almost certainly undetectable with the current generation of X-ray telescopes. 

\begin{acknowledgements}
We thank Dr.\ B.\ Verghese for help in preparation of the manuscript.
SR thanks IIA and PJW thanks TIFR for hospitality during this work.
PJW is grateful for hospitality from the Astrophysical Sciences Department
at Princeton University and for support from Research Program Enhancement
funds at GSU.
\end{acknowledgements}


\begin{thebibliography}{}
\bibitem[1996]{abramowicz} Abramowicz, M.\ A., Chen, X.-M., Granath, M., \& Lasota, J.-P.
1996, ApJ, 471, 762
\bibitem[1982]{antonucci} Antonucci, E., Gabriel, A.\ H., Acton, L.\ W., 
et al. 1982, Sol.\ Phys.\ 78, 107
\bibitem[1998]{bastian} Bastian, T.\ S., Benz, A.\ O., \& Gary, D.\ E. 1998,
ARA\&A, 36, 131
\bibitem[1998]{brandenburg} Brandenburg, A. 1998, Disc Turbulence and Viscosity, 
in Theory of Black Hole Accretion Disks,
ed.\ M.\ A.\ Abramowicz, G.\ Bj{\"o}rnsson,  \& J.\ E.\ Pringle 
(Cambridge U.\ Press, Cambridge), 61
\bibitem[1994]{chakrabarti94} Chakrabarti, S.\ K., \& D'Silva, S. 1994, ApJ, 424, 138
\bibitem[1995]{chakrabarti95} Chakrabarti, S.\ K., \& Titarchuk, L.\ G. 1995, 
ApJ, 455, 623
\bibitem[1998]{dimatteo} di Matteo, T. 1998, MNRAS, 299, L15
\bibitem[1997]{esin97} Esin, A.\ A., McClintock, J.\ E., \& Narayan, R.
1997, ApJ, 489, 865 
\bibitem[1998]{esin98} Esin, A.\ A., Narayan, R., Cui, W., Grove, J.\ E.,
\& Zhang, S.-N. 1998, ApJ, 505, 854
\bibitem[1987]{fisher} Fisher, G.\ H. 1987, ApJ, 317, 502
\bibitem[1985]{fisheretal} Fisher, G.\ H., Canfield, R.\ C., \& McClymont, A.\ N., 1985, ApJ, 289,
414
\bibitem[1979]{galeev} Galeev, A.\ A., Rosner, R., \& Vaiana, G.\ S. 1979, ApJ, 229, 318
\bibitem[1994]{haardt} Haardt, F., Maraschi, L., \& Ghisellini, G. 1994, ApJ, 432, L95
\bibitem[1996]{hawley} Hawley, J.\ F.,  Gammie, C.\ F., \& Balbus, S.\ A.
1996, ApJ, 464, 690
\bibitem[1994]{krishan94} Krishan, V., \& Wiita, P.\ J. 1994, ApJ, 423, 172
\bibitem[2000]{krishan00} Krishan, V., Wiita, P.\ J., \& Ramadurai, S. 2000, A\&A, 356, 373
\bibitem[1995]{kuijpers} Kuijpers, J. 1995, Flares in Accretion Disks, 
in Coronal Magnetic Energy Releases, ed.
A.\ O.\ Benz  \& A.\ Kr{\"u}ger (Springer, Berlin), 135
\bibitem[2000]{lakhina} Lakhina, G.\ S. 2000, Bull.\ Astr.\ Soc.\ India, 28, 593
\bibitem[1996]{lasota} Lasota, J.-P., Abramowicz, M. A., Chen, X.,
 Krolik, J.,  Narayan, R., \& Yi, I. 1996, ApJ, 462, 142L
\bibitem[1993]{lawrence} Lawrence, A., \& Papadakis, I. 1993, ApJ, 414, L85
\bibitem[1986]{mcclymont} McClymont, A.\ N., \& Canfield, R.\ C. 1986, ApJ, 305, 936
\bibitem[2000]{miller} Miller, K.\ A., \& Stone, J.\ M. 2000, ApJ, 534, 398
\bibitem[1997]{narayan97} Narayan, R., Kato, S., \& Honma, F. 1997, ApJ, 476, 49
\bibitem[1996]{narayan96} Narayan, R., Yi, I., \& Mahadevan, R. 1996, A\&AS, 120, 287
\bibitem[1988]{parker} Parker, E.N. 1988, ApJ, 330, 474
\bibitem[1998]{poutanen} Poutanen, J. 1998, Accretion disc-corona models 
and X/$\gamma$-ray spectra of accreting black holes,
in Theory of Black Hole Accretion Disks,
ed.\ M.\ A.\ Abramowicz, G.\ Bj{\"o}rnsson,  \& J.\ E.\ Pringle 
(Cambridge U.\ Press, Cambridge), 100
\bibitem[1976]{raymond} Raymond, J.\ C., Cox, D.\ P., \& Smith, B.\ W. 1976, ApJ, 204, 290
\bibitem[2000]{robinson} Robinson, P.\ A., \& Benz, A.\ O. 2000, Solar Phys., 194, 345
\bibitem[1973]{shakura} Shakura, N.\ I., \& Sunyaev, R.\ A., 1973, A\&A, 24, 337 (SS)
\bibitem[1996]{tanaka} Tanaka, Y., \& Shibazaki, N. 1996, ARA\&A, 34, 607
\bibitem[2001]{trudolyubov}  Trudolyubov, S.\ P., Borozdin, K.\ N., \& Priedhorsky, W.\ C. 2001,
MNRAS, 322, 309
\bibitem[1999]{turner} Turner, T.\ J., George, I.\ M., Nandra, K., \&
Turcan, D. 1999, ApJ, 524, 667 
\bibitem[1995]{wagner}  Wagner, S.\ J., \& Witzel, A. 1995, ARA\&A, 33, 163
\bibitem[1982]{white}  White, N.\ E., \& Holt, S.\ S. 1982, ApJ, 257, 318
\bibitem[1996]{wiita}  Wiita, P.\ J. 1996, Accretion Disk Models for Rapid
Variability, in Blazar Continuum Variability, ed.\ H.\ R.\ Miller, 
J.\ Webb, \& J.\ C.\ Noble  (ASP, San Francisco), 42
\end{thebibliography}
\end{document}